\documentclass[sigconf]{acmart}

\usepackage[inline]{enumitem}
\usepackage{subcaption}
\usepackage{multirow}

\usepackage{booktabs}
\usepackage{amsmath}
\usepackage{tablefootnote}

\usepackage{wrapfig}

\usepackage{xcolor} % remove this for submission
\AtBeginDocument{%
  }

\copyrightyear{2024}
\acmYear{2024}
\setcopyright{rightsretained}
\acmConference[HAI '24]{International Conference on Human-Agent Interaction}{November 24--27, 2024}{Swansea, United Kingdom}
\acmBooktitle{International Conference on Human-Agent Interaction (HAI '24), November 24--27, 2024, Swansea, United Kingdom}\acmDOI{10.1145/3687272.3688329}
\acmISBN{979-8-4007-1178-7/24/11}

\begin{document}

\title{Combating Spatial Disorientation in a Dynamic Self-Stabilization Task Using AI Assistants}

\author{Sheikh Mannan}
\email{sheikh.mannan@colostate.edu}
\affiliation{%
  \institution{Colorado State University}
  \city{Fort Collins}
  \state{CO}
  \country{USA}
}

\author{Paige Hansen}
\email{p.hansen@colostate.edu}
\affiliation{%
  \institution{Colorado State University}
  \city{Fort Collins}
  \state{CO}
  \country{USA}
}

\author{Vivekanand Pandey Vimal}
\email{somde@brandeis.edu}
\affiliation{%
  \institution{Brandeis University}
  \city{Waltham}
  \state{MA}
  \country{USA}
}

\author{Hannah N. Davies$^*$}
\email{hannahndavies@gmail.com}
\affiliation{%
  \institution{Emory University School of Medicine}
  \city{Atlanta}
  \state{GA}
  \country{USA}
}

\author{Paul DiZio}
\email{dizio@brandeis.edu}
\affiliation{%
  \institution{Brandeis University}
  \city{Waltham}
  \state{MA}
  \country{USA}
}

\author{Nikhil Krishnaswamy}
\email{nkrishna@colostate.edu}
\affiliation{%
  \institution{Colorado State University}
  \city{Fort Collins}
  \state{CO}
  \country{USA}
}

\renewcommand{\shortauthors}{Mannan et al.}


% As a minor point, valuable data were not described in the abstract. It would have been easier for the reader to understand if representative data had been presented, even if briefly.
\begin{abstract}
  Spatial disorientation is a leading cause of fatal aircraft accidents. This paper explores the potential of AI agents to aid pilots in maintaining balance and preventing unrecoverable losses of control by offering cues and corrective measures that ameliorate spatial disorientation. A multi-axis rotation system (MARS) was used to gather data from human subjects self-balancing in a spaceflight analog condition. We trained models over this data to create ``digital twins'' that exemplified performance characteristics of humans with different proficiency levels. We then trained various reinforcement learning and deep learning models to offer corrective cues if loss of control is predicted. Digital twins and assistant models then co-performed a virtual inverted pendulum (VIP) programmed with identical physics. From these simulations, we picked the 5 best-performing assistants based on task metrics such as crash frequency and mean distance from the direction of balance. These were used in a co-performance study with 20 new human subjects performing a version of the VIP task with degraded spatial information. We show that certain AI assistants were able to improve human performance and that reinforcement-learning based assistants were objectively more effective but rated as less trusted and preferable by humans.
\end{abstract}

\begin{CCSXML}
<ccs2012>
   <concept>
       <concept_id>10003120.10003121.10003124</concept_id>
       <concept_desc>Human-centered computing~Interaction paradigms</concept_desc>
       <concept_significance>500</concept_significance>
       </concept>
   <concept>
       <concept_id>10003120.10003121.10011748</concept_id>
       <concept_desc>Human-centered computing~Empirical studies in HCI</concept_desc>
       <concept_significance>500</concept_significance>
       </concept>
   <concept>
       <concept_id>10010147.10010178.10010213</concept_id>
       <concept_desc>Computing methodologies~Control methods</concept_desc>
       <concept_significance>300</concept_significance>
       </concept>
   <concept>
       <concept_id>10010147.10010257.10010258</concept_id>
       <concept_desc>Computing methodologies~Learning paradigms</concept_desc>
       <concept_significance>300</concept_significance>
       </concept>
 </ccs2012>
\end{CCSXML}

\ccsdesc[500]{Human-centered computing~Interaction paradigms}
\ccsdesc[500]{Human-centered computing~Empirical studies in HCI}
\ccsdesc[300]{Computing methodologies~Control methods}
\ccsdesc[300]{Computing methodologies~Learning paradigms}

\keywords{Spatial disorientation, Balancing, AI assistance, Real-time human-agent interaction}


\maketitle
\def\thefootnote{*}\footnotetext{This work performed at Brandeis University.}\def\thefootnote{\arabic{footnote}}

\vspace*{-2mm}
\section{Introduction}

Maintaining spatial awareness and orientation is critical in domains like piloting, spaceflight, and even driving. Spatial disorientation has been and continues to be a leading cause of fatal aircraft incidents~\cite{braithwaite1998spatial,gibb2011spatial} and occurs when sensory information (e.g. from the visual, somatosensory, and vestibular systems) is erroneous, which can lead to unrecoverable crashes, injury, or loss of life~\cite{newman2007overview,gibb2011spatial}. 

An AI agent in this situation could potentially use numerical signals to track the pilot and vehicle's positioning in the relevant orientational plane(s), detect if there is a risk of losing control~\cite{daiker2018use,zgonnikova2016stick,wang2022crash}, and even alert the pilot to make corrective maneuvers. However, there is a record, particularly in high-risk domains like aviation, of either over-trust or under-trust in highly automated systems. For instance, \citeauthor{sadler2016effects} \cite{sadler2016effects} demonstrated that pilots' trust in the recommendations of an automated system correlated with the level of transparency (such as justification) in the recommendation. The {\it shared autonomy} literature indicates that even when an agent knows an optimal strategy, failing to comply with a suboptimal strategy its human partner insists on may have a negative effect on trust and lead to disuse of the system \cite{hancock2011meta,lee2013computationally,salem2015would,nikolaidis2017human}. 

In this paper, we hypothesize that when attempting to balance themselves under disorienting conditions, humans will be more receptive to assistance from an AI whose strategy to regain balance is more human-like, even if that strategy is objectively less optimal than a less human-like strategy. We establish a novel task of AI assistance in regaining balance in disorienting situations using a documented, realistic simulation of vehicle control in a spaceflight analog condition. In this scenario, subjects are deprived of gravitational cues and use a joystick to self-balance while seated in a {\it multi-axis rotation system} (MARS) programmed to behave like an inverted pendulum \cite{panic2015direction,vimal2017learning,vimal2019learning}. We took data of human subjects attempting to keep the MARS balanced and used it to train digital twins of ``pilots'' that exemplified performance characteristics of humans of various proficiency levels.  We then trained multiple ``assistant'' models that, due to different data and techniques, demonstrate performance strategies that may align with or differ drastically from those of humans.
%{\it embody} the problem space in ways that may or may not align with humans.
The pilot and assistant models were placed in a co-performance simulation with a {\it virtual inverted pendulum} (VIP) programmed with identical physics to the MARS. Assistants attempted to help the pilot models keep the VIP balanced and avert crashes by offering corrective cues when the pilot risked a destabilizing loss of balance. To test transfer to a different environment, we ran assistants with digital twins trained over data from humans performing an instance of the VIP task directly. Finally, the best-performing assistant models were deployed in a human-subject study in the VIP environment to demonstrate the feasibility of AI assistance as a countermeasure in disorienting situations.
%We assess how much each type of assistant helped different pilot models improve.
%and characterize assisted performance with respect to the level of learning displayed by real human participants in the MARS and VIP tasks

Following a discussion of related literature (Sec.~\ref{sec:related}), in Sec.~\ref{sec:tasks} we describe the MARS and VIP tasks, the human performance data collected from each, and how performance is evaluated in these tasks. Sec.~\ref{sec:training} describes the training of different reinforcement learning models using an analogous environment to the MARS/VIP tasks, and deep learning models using human performance data, as well as the selection of ``digital twin" models of different human proficiencies. Sec.~\ref{sec:eval} presents the evaluation methods and results for two studies: the high-throughput digital twins study wherein 21 candidate assistant models made corrective suggestions to the different digital twins of human pilots, and a human-subject study wherein the five best performing assistants from the digital twins study engaged in co-performance and co-training in the VIP task with real humans. In Sec.~\ref{sec:disc} and Sec.~\ref{sec:conc}, we discuss our findings and their implications for human-agent collaboration and trust when performing real-time situated tasks, and directions for future work.\footnote{Our code and data is available at \url{https://github.com/csu-signal/HITL-VIP}.}

Our novel contributions are:
\begin{enumerate*}[label=\arabic*)]
    \item a novel task of AI assistance in disoriented self-balancing, a challenging action-learning task with well-controlled parameters;
    \item an assessment of different reinforcement learning and deep learning models' abilities to prevent destabilizing loss of control in this task using a high-throughput digital twins setting;
    %\item an analysis of the properties of assistant models to inform the construction and deployment of successful assistants in co-performance with real humans.
    \item a human-agent co-performance study with the best-performing AI assistants involving human-in-the-loop (HITL) AI training, to demonstrate transfer to real humans.
\end{enumerate*}

\vspace*{-2mm}
\section{Related Work}
\label{sec:related}

\paragraph{Spatial disorientation and balance}
An inverted pendulum (with center of mass above the pivot point) is a common model of human upright balance in the study of postural dynamics \cite{riccio1992role}. \citeauthor{panic2015direction} \shortcite{panic2015direction} and \citeauthor{vimal2016learning} \shortcite{vimal2016learning} explored the relevance of the MARS balancing task to the perception of gravitational cues during unstable vehicle control. Subjects were strapped into a MARS device programmed with inverted pendulum (IP) dynamics and instructed to stabilize themselves about the direction of balance (DOB) using a joystick. Because the risk of spatial disorientation-related accidents is heightened when visual information is limited \cite{lyons2006aircraft,takada2009survey}, subjects were blindfolded. Typically, humans rely on gravitational cues when balancing, which are detected by the vestibular and somatosensory systems as participants tilt away from the gravitational vertical, however in spaceflight conditions gravitational cues are not reliable. To create a disorienting spaceflight analog condition, \citeauthor{panic2017gravitational} \shortcite{panic2017gravitational} and \citeauthor{vimal2017learning} \shortcite{vimal2017learning,vimal2019learning} placed participants in the Horizontal Roll Plane, where they were always perpendicular to the gravitational vertical and no longer tilting relative to it. 90\% of participants reported spatial disorientation and in data 100\% exhibited characteristic positional drifting \cite{vimal2019learning}. Participants showed minimal learning and frequent ``crashes'' (reaching pre-programmed $\pm$60$^\circ$ boundaries, after which the MARS automatically reset to the DOB).

\paragraph{AI algorithms} Tasks like IP balancing are well-known use cases in reinforcement learning \cite{barto1983neuronlike,anderson1989learning,florian2007correct} that serve as demonstration benchmarks for newer continuous control algorithms like SAC \cite{haarnoja2018soft} or DDPG \cite{lillicrap2015continuous}. These have demonstrated proficiency at solving non-linear control tasks like IP balancing using reward signals extracted from observation of applied environmental physics. However, due to the nature of environmental physics input vs. sensorimotor input, we hypothesize that they learn to perform the task very differently from humans. Here we explore the application of different RL and deep learning algorithms to AI assistance in disoriented balancing.\footnote{Although the RL approaches we use are also parameterized by multilayer neural networks, we use ``deep learning'' to refer to non-reinforcement learning algorithms, such as sliding window or sequence modeling approaches.}

\paragraph{Embodiment} 

Seminal literature operationalizes embodiment as a two-way process between brain and body/environment \cite{thompson2001radical,varela2017embodied}. Most modern approaches treat embodiment primarily at a surface level, such as the richness of visual \cite{arzy2006neural-mech,arzy2006neural-basis,hellmann2011effect,hellman2015robot} or interoceptive \cite{allen2020thinking} representation of an agent’s form, or in terms of the physical form an agent takes, wherein the actions it is capable of are conditioned upon its articulators \cite{pfeifer2001understanding,di2014enactive,pustejovsky2021embodied}. We adopt a definition of embodiment akin to Ziemke’s \textit{structural coupling} \cite{ziemke2001robots}, where a system embodied in an environment takes environmental states as input, which changes the condition of the system \cite{quick1999making}. However, the types of information a system is exposed to, such as through different types of sensors, also condition the relations the system develops between itself and its environment \cite{cariani1992some}, such that exposure to different types of data (or different ways of measuring the same underlying environmental state) may mean that different inputs and modeling strategies cause a model to learn different policies within the same action space, and thus may learn to perform the same task equally well through potentially radically different strategies. Therefore we will speak of different ``embodiments'' of the task problem space as reflected through these different strategies.

%is often thought of in terms of the physical form an agent takes, wherein the actions it is capable of are conditioned upon the articulators it possesses \cite{pfeifer2001understanding,di2014enactive,pustejovsky2021embodied}. However, the types of information a system is exposed to, such as through different types of sensors, also condition the relations the system develops between itself and its environment \cite{cariani1992some}. Using machine learning to find patterns in data is one way of modeling such relations, but exposure to different types of data may mean that different models learn to perform the same task in different ways. We adopt a definition of embodiment akin to Zimekie \shortcite{ziemke2001robots}'s {\it structural coupling}, where a system embodied in an environment takes environmental states as input, which changes the condition of the system \cite{quick1999making}. Differences in how environmental states are measured also result in different outputs or changes to the AI system. \todo{Embodiment: different inputs cause a model to learn different policies within the same action space} 

\begin{figure}[t]
\centering
\includegraphics[width=\columnwidth]{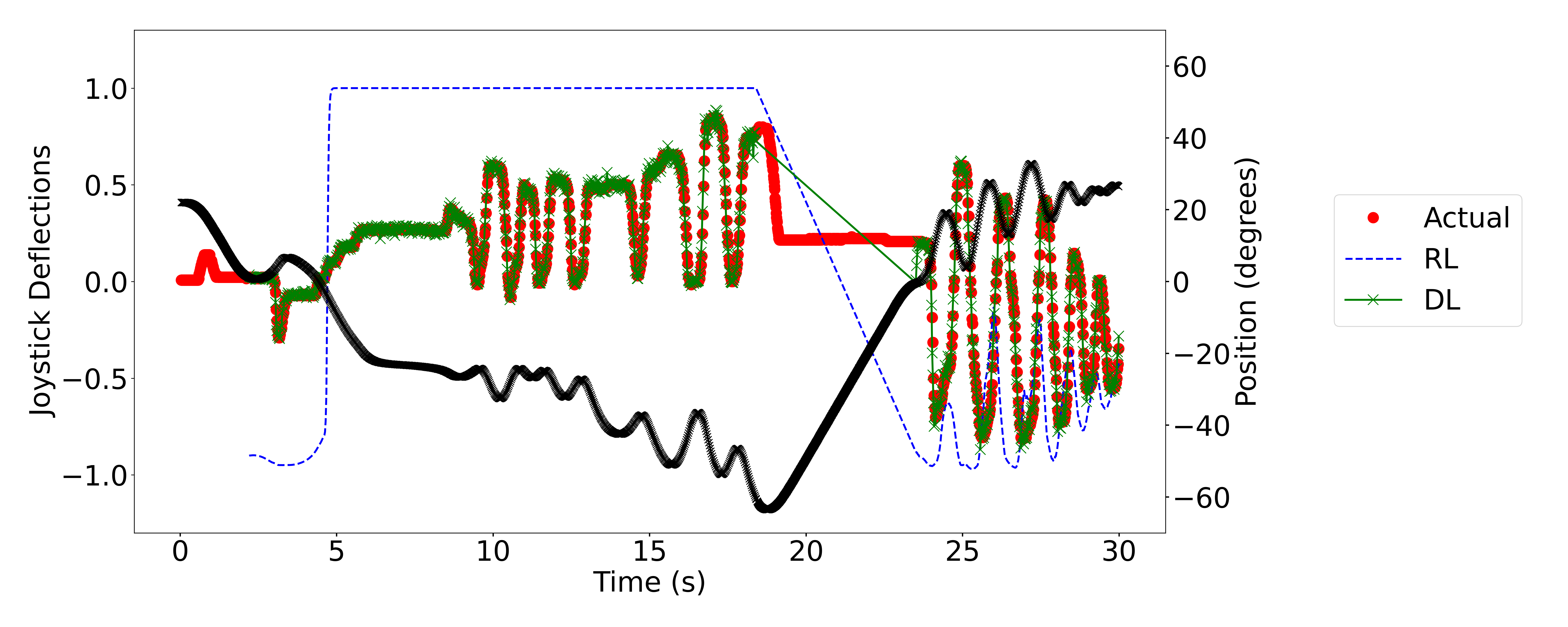}
\caption{Joystick deflections predicted by a DDPG (blue) and an LSTM trained over human data (green) compared to an actual 30-sec. MARS participant trial sample (participant deflections in red and angular position in black). This instance of the LSTM displays a test RMSE of .013 while the DDPG gets .803.}
\vspace*{-2mm}
\label{fig:defl}
\end{figure}

Fig.~\ref{fig:defl} shows how this manifests in the MARS task. ``Actual'' shows the deflections of a human subject---the subject balanced the MARS using small, intermittent deflections that are characteristic of proficient human performance~\cite{vimal2020characterizing}. Joystick deflections predicted by a DDPG model are shown in blue and deflections predicted by an LSTM trained over actual human motions from other MARS trials are shown in green. The model trained on human data \textit{embodies} the problem space similarly to a proficient human, making it a superior predictor of novel humans' small, intermittent actions, while the DDPG predicts long, continuous deflections that are more indicative of poor human performance, even though a DDPG successfully performs the task independently (see also \cite{mannan2024embodying}). In this paper, we assess if and how an AI model's ability to embody a task space similarly to a human translates to increased ability to assist a human in the task or greater human preference for that assistant compared to others.

\paragraph{Trust}
%\todo{Abridge Shneiderman discussion, move Palmer to Discussion section with Human-AI results}
% Trust, I better read the AI trust lit this week

% \todo{For each of these sections, how does this work extend the prior related work?}

% \todo{\citet{doi:10.1080/24721840.2022.2127724} may be a good resource but I could not find an accessible link to it - mannan}

%Shneiderman \shortcite{shneiderman2020humancentered}'s human-centered AI (HCAI) framework separates levels of autonomy and levels of human control in pursuit of systems that achieve high levels of both, such that the human feels in control of task execution at all times, but also achieves better performance with the presence of increased automation than without it. 

Life-critical situations involving dynamic systems, such aviation use cases, may require rapid decisions that may have irreversible consequences. According to Schneiderman \shortcite{shneiderman2020humancentered}'s human-centered AI (HCAI) framework, systems deploying high levels of autonomy would need to be \textit{reliable}, \textit{safe}, and \textit{trustworthy} (RST) within known parameters. Human users must also not become over-reliant on automation such that if the system makes a mistake, the human can still override the AI using situational knowledge and value judgments. Furthermore, automated systems may prompt over-reliance in emergency scenarios even after the systems have demonstrated a lack of reliability \cite{robinette2016overtrust,wagner2018overtrust,tomzcak2019let}, or an under-reliance on a typically robust system after a single failure point \cite{parasuraman1997humans}.

% {\todo{Over-reliance} Furthermore, there are gray areas, excessive human control, or automation that need to be avoided. In order to avoid losing trust in AI systems after a mishap or accident, the level of autonomy provided needs to be explained to prevent humans from being suggested there is greater capability than is available. 
% }

% \sm{
% \citet{Palmer2016} illustrate trust dimensions that arise from the use of an autonomous or automated system(s) such as: 
% \begin{itemize}
%     \item Reliability - the system has a small chance of failure - what areas do our assistants fail to perform?
%     \item Robustness - the system can handle perturbations/deviations appropriately - when noise is added to the action, how well does it still perform?
%     \item Understandability - the conclusions reached or actions taken by the system can be understood - do the assistant actions make sense?
%     \item Benevolence—system is supporting the mission and operator (and not in opposition) - does the assistant help or fight the pilot and vice versa?
%     \item Adaptive/Learning—system can acquire information and then beneficially leverage that information - can the assistant improve by learning new information?
%     \item Adversarial—system can complete its mission when subject to opposing efforts - ??
%     \item Dynamic—system can complete its mission in a changing environment - can the assistant trained on MARS data help pilots in the VIP environment? 
% \end{itemize}
% }

\vspace*{-2mm}
\section{MARS and VIP Tasks}
\label{sec:tasks}
\begin{figure}[t]
\centering
\includegraphics[width=0.95\columnwidth]{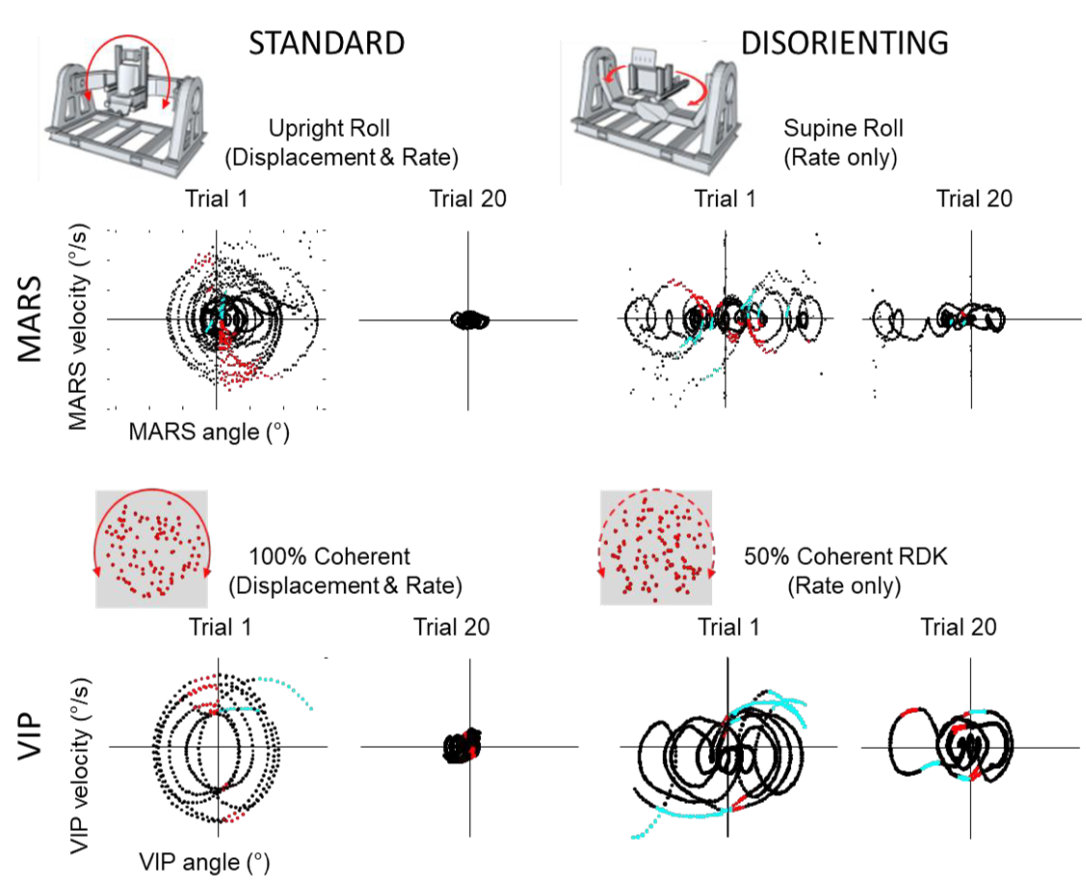} % Reduce the figure size so that it is slightly narrower than the column. Don't use precise values for figure width.This setup will avoid overfull boxes.
\vspace*{-2mm}
\caption{Typical performance in the MARS and VIP tasks, before practice (Trial 1) and after practice (Trial 20). Phase plots show angular velocity vs. angular displacement from the DOB. The ``standard'' conditions provide angular displacement and velocity cues, and subjects improve significantly between first and last trial, seen as clustering around the origin (balance point) by Trial 20. The ``disorienting'' conditions eliminate sensory signals about displacement from the DOB, increasing positional drift (shown as phase loop oscillations around the X-axis) and destabilizing joystick commands that accelerate away from the DOB in the current direction of motion, with minimized learning and continued positional drift in Trial 20. Cyan dots indicate destabilizing deflections, where position, velocity and joystick deflection all have the same sign. Red dots denote {\it anticipatory} deflections, where position and joystick deflection have the same sign but velocity has the opposite sign---usually done to slow the IP down when velocity is perceived as being too high.}
\vspace*{-2mm}
\label{fig:mars-vip}
\end{figure}

%\todo{AAAI reviewer comment: The sections describing MARS and VIP tasks are hard to follow due to a lack of structure and excessive use of abbreviations. I recommend structuring this into subsections detailing 1) what these tasks are, 2) the performance metrics used, and 3) how the dataset was collected.}

Fig.~\ref{fig:mars-vip} compares the MARS and VIP paradigms. Both can be configured in challenging but non-disorienting modes, with standard sensory information; or difficult and disorienting modes, with degraded information.  In both disorienting conditions, subjects show the same characteristic drifting and lack of learning when compared to the coherent conditions.

The \textbf{MARS task} is described in Sec.~\ref{sec:related}. In the data we use here (described below), MARS dynamics were governed by $\ddot{\theta} = k_{P}sin\theta$, where $\theta$ is degrees deviation from the DOB and pendulum constant $k_{P} =$ 600$^{\circ}$/s$^2$. 

In the \textbf{virtual inverted pendulum (VIP)} paradigm, an analog to the MARS programmed with the same physics, subjects balance a visually simulated circular array of dots (random dot kinematogram, RDK) which rolls in the plane of the display screen. This is visually rendered for humans and can be directly actuated by an algorithm. In the disorienting VIP condition (similar to the Horizontal Roll Plane in the MARS), the RDK is 50\% coherent: alternating subsets of dots displace coherently across consecutive frames while the other half jump randomly. This eliminates configural displacement cues relative to the upright DOB while providing low-level retinal motion cues.  Similar performance degradations occur between MARS upright vs. supine and VIP 100\% vs. 50\% coherence conditions, even with practice, exhibited by an increased number of crashes accompanied by more frequent destabilizing actions (Fig.~\ref{fig:mars-vip}).

\begin{figure}[h!]
%\begin{wrapfigure}{l}{.45\columnwidth}
    \vspace*{-2mm}
    \centering
    \includegraphics[width=.5\columnwidth]{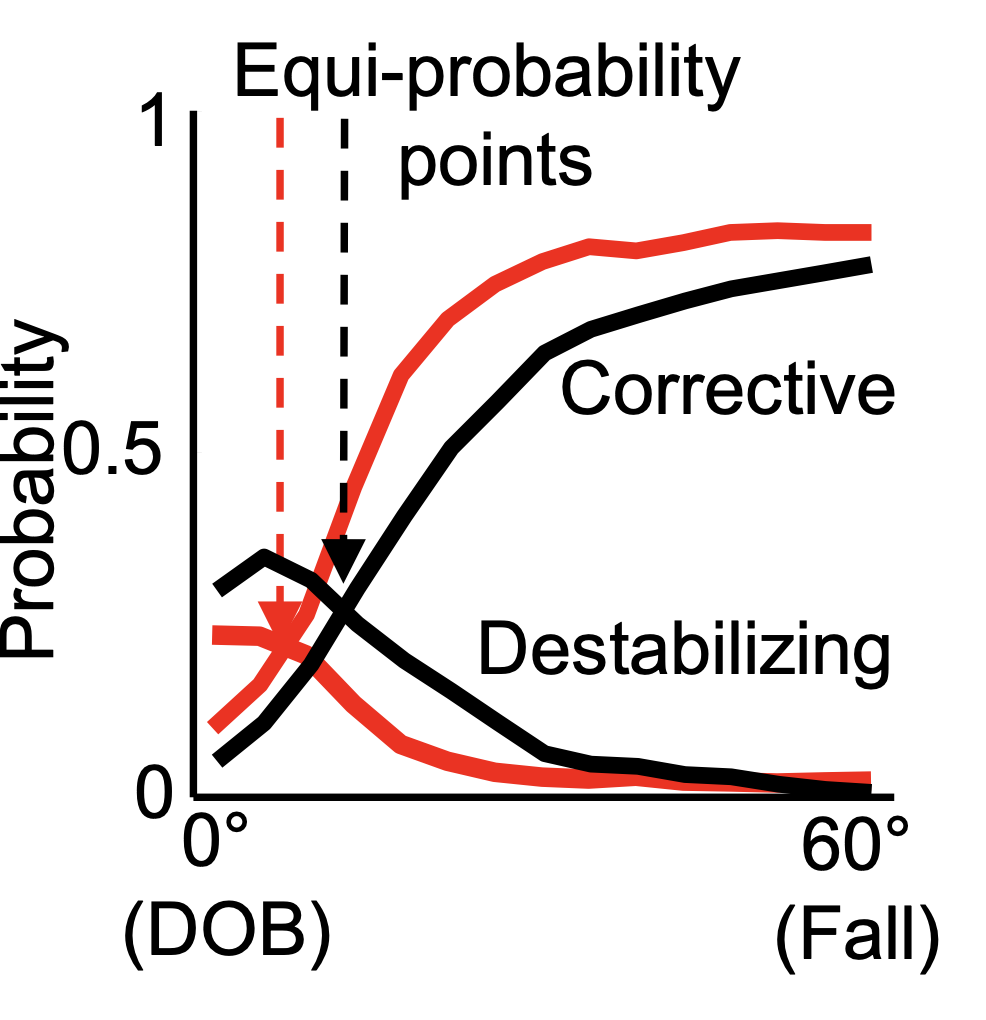}
    \vspace*{-2mm}
    \caption{Complementary evolution of discrete destabilizing and corrective commands as a function of angular deviation away from the DOB and toward a fall boundary, seen in MARS (red) and VIP (black) tasks.}
    \label{fig:equiprobability}
    \vspace*{-6mm}
%\end{wrapfigure}
\end{figure}

The same performance characteristics are exemplified in both MARS \& VIP, as defined by subject matter experts in the neuroscience of balance dynamics \shortcite{vimal2020characterizing}. In spaceflight, the pilot has cues about their motion but no orientation to gravity.  The VIP 50\% coherent reflects this by providing motion cues without cues about configural orientation. The VIP \& MARS tasks possess similar underlying physical models of instability and in both tasks, the probability of destabilizing commands decreases and corrective commands increase as the pendulum position crosses the DOB and moves closer to a fall boundary (Fig.~\ref{fig:equiprobability}). This shows how the VIP task is demonstrably analogous to the MARS task, as the MARS task is demonstrably analogous to spaceflight conditions. Importantly, these distributions of actions come from human task performance and demonstrate a particular kind of human action policy based on sensorimotor perception, an example of structural coupling.

\vspace*{-2mm}
\subsection{Data Collection}
\paragraph{MARS}
Wang et al. \shortcite{wang2022crash} released MARS human performance data from 34 healthy adults (18 female, 16 male). Each subject experienced two experimental sessions on consecutive days, each consisting of 20 100-sec. trials where they attempted to balance themselves with minimal oscillations while blindfolded. The data contains angular positions and velocities, and joystick deflections sampled at 50 Hz. 

\paragraph{VIP}
The VIP data consists of 31 healthy adults (22 female, 9 male). Subjects took part in 12 30-sec. trials in one session in the disorienting condition, with the same goal as the MARS task. Angular pos./vel. and joystick deflections were sampled at 200 Hz.

\vspace*{-2mm}
\subsection{Performance Evaluation}
\label{ssec:perform}
An ideal performer in both of these tasks would be one that immediately rotates to the balance point and stays there with little to no motion. Calculable metrics from the collected data include number of crashes (excursions beyond $\pm$60$^\circ$), proportion of destabilizing deflections (\textit{\% destab.}---see Fig.~\ref{fig:mars-vip}), mean/standard deviation of angular position $\theta$, average magnitude of velocity ($\mu\lvert Mag\rvert_{vel}$), and root-mean-square (RMS) velocity. Lower metric values usually mean improvement, e.g., fewer crashes, more time spent near the DOB, less oscillation, slower motion, smaller deflections, etc.
 
\paragraph{MARS}
Vimal et al. \shortcite{vimal2020characterizing} used a Bayesian Gaussian Mixture method and the aforementioned features to cluster subjects into 3 statistically distinct groups that represent Proficient, Somewhat-Proficient, and Not-Proficient performance (hereafter {\it Good}, {\it Medium} and {\it Bad}). 
%This was based on features like standard deviation of angular position ($\sigma(\theta)$), number of crashes (excursions beyond $\pm$60$^\circ$), 
%short-term diffusion coefficient from the Stabilogram Diffusion Function (SDF; \citeauthor{collins1993open} \shortcite{collins1993open}), 
%and average magnitude of velocity ($\mu\lvert Mag\rvert_{vel}$), percentage of destabilizing joystick deflections (\textit{\%destab}---see Fig.~\ref{fig:mars-vip}), 
%\textit{\%Anticipatory} (percentage of anticipatory joystick deflections, where MARS position and joystick deflection have the same sign but the MARS velocity has the opposite sign---usually done to slow the MARS down when velocity is perceived as being too high), Drift Rate, and $D_L$ (SDF long term diffusion coefficient). 
We uses these clusters to split the MARS data into training subsets and to characterize digital twin performance in the task.

\paragraph{VIP}
Plotting VIP subjects' RMS velocity vs. crash frequency revealed a positive linear relationship ($r=.73$).
%SDF diffusion coefficient, and velocity total power vs. fall frequency (\# crashes/30 sec.), and found a strong correlation ($r=$ .730, $r=$ .814, $r=$ .740 respectively). 
Based on the number of crashes in the 12$^{\text{th}}$ trial and crash reduction between Trials 1 and 12 (i.e., final performance and overall improvement), we assigned participants relative rankings and divided them in tertiles to mirror the Good/Medium/Bad MARS classification.  VIP participants typically exhibited more crashes, destabilizing actions, and RMS velocity compared to MARS participants of the equivalent proficiency. These factors and differences in sample rate and environment allow us to test the transfer of AI assistants to novel digital twins.

\vspace*{-1mm}
\section{Model Training}
\label{sec:training}

Our goals in model training were both to train AI models capable of independently performing an IP balancing task parameterized with MARS physics and to create digital twins of humans that replicate different kinds of participant performance in the task. In some cases these two categories overlapped, leading to a testable hypothesis: that a model that performs well at the task may also be able to assist a ``pilot'' (real or simulated) in performing the task better.

Fig.~\ref{fig:model-input-output} shows an I/O schematic of all models. At time $t_x$, models take in a window containing the past $winSize$ seconds of angular positions and velocities and predict the joystick deflection made at time $t_{x + future}$. DL models over human data additionally take in the past $winSize$ joystick deflections made by the subject. If $winSize = 0.0$, the input consists of the values at $t_x$ only. If $future = 0.0$, the next joystick deflection is predicted.

\vspace*{-2mm}
\subsection{Reinforcement Learning Models}

We trained reinforcement learning-based models that learn directly from exposure to environmental physics using a custom variation of Gymnasium's classic-control Pendulum environment \cite{towers_gymnasium_2023}. This included \begin{enumerate*}[label=\arabic*)]
    \item a problem space bounded at $\pm$60$^\circ$ from the DOB, like the MARS/VIP task;
    \item a random starting point for the inverted pendulum within the newly defined problem space;
    \item a custom reward function\footnote{Where $\theta$ is the angular position, $\omega$ is the angular velocity, and $d$ is the joystick deflection.} given by Eq.~\ref{eq:rl}, to encourage small continuous adjustments like those of Good MARS participants \cite{vimal2019learning,vimal2020characterizing}. 
\end{enumerate*} 

\begin{equation}
            r=
    \begin{cases}
        \hfil 0, & \text{if } -30^\circ \leq \theta \leq 30^\circ \\
        -(\theta^2+.1\omega^2+.01d^2), & \text{if } \theta < 30^\circ \cup \theta > 30^\circ
    \end{cases}
    \label{eq:rl}
\end{equation}

The RL algorithms were directly exposed to environmental physics, unlike the DL models which received only implicit physics through the human performance data. The default SAC and DDPG implementations routinely converged to an optimal strategy that manifests as rotating immediately to the DOB and holding position there. We also trained and evaluated behavior cloning (BC)~\cite{ross2010efficient} and adversarial inverse RL (AIRL)~\cite{fu2018learning} using Good MARS participant data, to teach the models strategies closer to what humans would execute, in terms of replicating behavior or uncovering implicit reward functions in the data.

\begin{figure}[t]
\centering
\includegraphics[width=0.8\columnwidth]{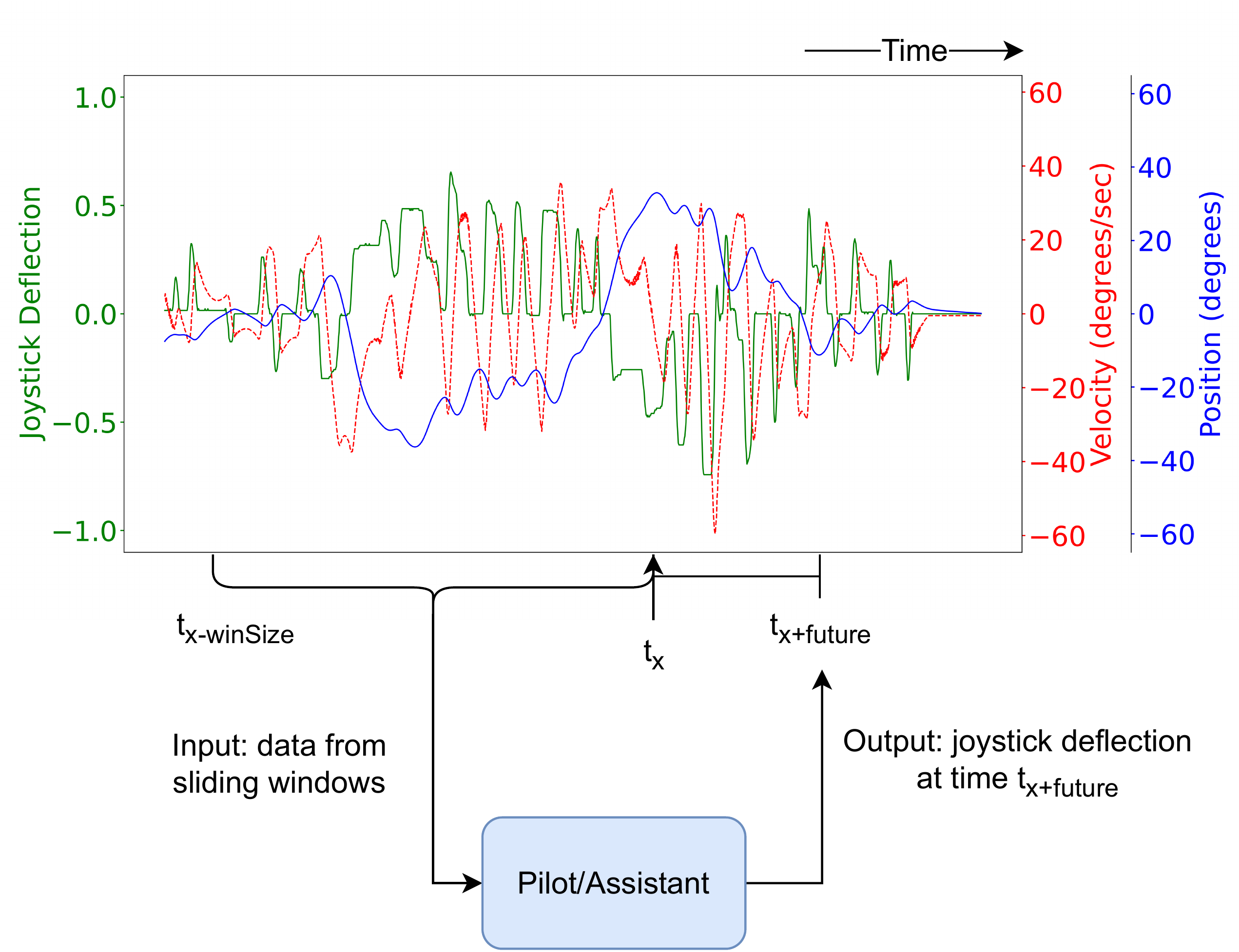} % Reduce the figure size so that it is slightly narrower than the column. Don't use precise values for figure width.This setup will avoid overfull boxes.
\vspace*{-2mm}
\caption{Model input and output structure. ``Pilot/Assistant'' stands in for any one of the trained prediction models.}
\vspace*{-4mm}
\label{fig:model-input-output}
\end{figure}

RL models take the current angular position and velocity to predict the next joystick deflection ($winSize=0.0,future=0.0$, cf. Fig.~\ref{fig:model-input-output}). We used Stable-Baselines3's SAC and DDPG implementations \cite{raffin2021stable}, and trained them with the default MLP policy, BC, or AIRL. Gaussian distributed noise was added to the action space to encourage exploration as the IP is considered an under-actuated task \cite{hollenstein2022action}. We trained 5 RL models:
\begin{enumerate*}[label=\arabic*)]
    \item SAC \& DDPG each with the standard policy;
    \item SAC \& DDPG each trained using BC;
    \item AIRL implemented with a SAC-based generator model. 
\end{enumerate*} %Hyperparameter details are in the appendix.

\vspace*{-2mm}
\subsection{Deep Learning Models}

To replicate human-like real-time performance of the MARS task, we trained Multilayer Perceptron (MLP), Vanilla Recurrent (RNN), Long-Short Term Memory (LSTM) \shortcite{hochreiter1997long}, and Gated Recurrent Unit (GRU) \shortcite{cho2014learning} network architectures over actions made by humans in the actual MARS data. 
%Input at time $t_{x}$ is the current angular position and velocity of the MARS and current joystick deflection made by the human, plus a window containing $winSize$ past angular positions, velocities, and joystick deflections. A successful model predicts a joystick deflection made at time $t_{x+future}$. Fig.~\ref{fig:model-input-output} shows a general schematic. 
Architectures were trained using different window sizes ($0.0s$, just the current timestep---MLP models only; and $0.2s$, $0.3s$, and $0.5s$).  Training data was also split into Good, Medium, and Bad proficiencies and individual models were trained on data of a specific proficiency. An additional set of models was trained using a combination of 1) Good \& Medium and 2) Good, Medium \& Bad proficiency data, to see if models could learn strategies employed by certain proficiency groups in scenarios that were not experienced by the others. In total, we trained 40 individual DL models, all of which, even when they successfully avoid crashes in solo task performance, demonstrate suboptimal strategies with human-like oscillation and intermittent deflections. These behavioral differences show the differences in how the RL and DL models learn to situate themselves in (or ``embody'') the problem space. %See the appendix for the features of each, and hyperparameter details.

% \begin{table*}
% \centering
% \begin{tabular}{lcccccccc}
% \toprule
%         {\bf Pilot} & {\bf Crashes}  & {\bf \% destab.}   & {\bf $\mu\lvert\theta\rvert$ ($^\circ$)} & {\bf $\sigma(\theta)$ ($^\circ$)} & {\bf $\mu\lvert Mag\rvert_{vel}$ ($^\circ$/s)} & {\bf $\sigma(\lvert Mag\rvert_{vel})$ ($^\circ$/s)} & {\bf $vel$ RMS}         & {\bf $\mu\lvert d\rvert$} \\
% \midrule
% Good    & \small 7 / 17  & \small 15.9 / 54.0 & \small 16.6 / 20.3  & \small 21.5 / 23.7    & \small 53.3 / 53.0    & \small 69.1 / 73.3    & \small 70.6 / 77.7    & \small .24 / .11  \\
% Med.    & \small 9 / 40  & \small 21.4 / 63.3 & \small 20.1 / 28.8  & \small 18.9 / 29.7    & \small 68.3 / 122.7   & \small 89.3 / 149.4   & \small 93.9 / 152.3   & \small .36 / .77  \\
% Bad     & \small 27 / 23 & \small 36.7 / 52.8 & \small 21.4 / 19.4  & \small 25.9 / 14.3    & \small 114.1 / 57.4   & \small 121.9 / 72.3   & \small 135.9 / 92.2   & \small .50 / .18  \\  
% \bottomrule
% \end{tabular}
% \caption{Performance statistics of pilot exemplar models (values are averaged over 3$\times$30 sec. trials except \# crashes, which is summed). Slashes separate models trained over MARS and VIP data. Columns from L--R: \# crashes, \% destabilizing actions, mean and SD distance from DOB, mean and SD angular velocity magnitude, RMS velocity, and mean deflection magnitude.}
% \label{tab:pilots}
% \end{table*}

\begin{table*}
\centering
\begin{tabular}{lccccccc}
\toprule
        {\bf Pilot} & {\bf Crashes}$\downarrow$  & {\bf \% destab.}$\downarrow$   & {\bf $\mu\lvert\theta\rvert$ ($^\circ$)}$\downarrow$ & {\bf $\sigma(\theta)$ ($^\circ$)}$\downarrow$ & {\bf $\mu\lvert Mag\rvert_{vel}$ ($^\circ$/s)}$\downarrow$ & {\bf $vel$ RMS}$\downarrow$  \\
\midrule
\small Good    & \small 7 / 17  & \small 15.9 / 54.0 & \small 16.6 / 20.3  & \small 21.5 / 23.7    & \small 53.3 / 53.0    & \small 70.6 / 77.7   \\
\small Med.    & \small 9 / 40  & \small 21.4 / 63.3 & \small 20.1 / 28.8  & \small 18.9 / 29.7    & \small 68.3 / 122.7   & \small 93.9 / 152.3  \\
\small Bad     & \small 27 / 23 & \small 36.7 / 52.8 & \small 21.4 / 19.4  & \small 25.9 / 14.3    & \small 114.1 / 57.4   & \small 135.9 / 92.2   \\  
\bottomrule
\end{tabular}
\caption{Performance statistics of pilot exemplar models (values are averaged over 3$\times$30 sec. trials except \# crashes, which is summed). Slashes separate models trained over MARS and VIP data. Columns from L--R: \# crashes, \% destabilizing actions, mean and SD distance from DOB, mean and SD angular velocity magnitude, and RMS velocity. Lower values are better (Sec.~\ref{ssec:perform}).}
\vspace*{-4mm}
\label{tab:pilots}
\end{table*}

\subsubsection{Selecting Representative Pilots}
\label{sssec:pilot-selection}

As the same performance characteristics are exemplified in both MARS \& VIP, we identified models that most closely approximate performance categories from \citeauthor{vimal2020characterizing} \shortcite{vimal2020characterizing}, reflected in Table~\ref{tab:pilots}.

All DL models were made to perform the VIP task ($3\times30s$ trials), with angular position, velocity, and joystick deflection recorded at each timestep. We extracted performance features shown in Table~\ref{tab:pilots}. Since the data distribution could not be assumed to be spherical, these features were used in $k$-means clustering ($k=3$) to approximate the split into Good, Medium, or Bad groups. Following \citeauthor{vimal2019learning} \shortcite{vimal2019learning}, the cluster of models that displayed higher oscillations and greater average magnitude of deflections was considered Bad while the cluster that displayed smaller, more intermittent actions was considered Good (with the remainder considered Medium). We then took the models in each cluster that were trained over the equivalent data subset (e.g., models in the Good cluster trained over Good data, m.m.), and used the VIP performance characterization technique from Sec.~\ref{ssec:perform} to identify which model best exemplified the characteristics of each proficiency group: \textbf{Good}---LSTM trained over Good data with a window size of 0.2s; \textbf{Medium}---GRU trained over Medium data with a window size of 0.3s predicting 0.1s into the future; \textbf{Bad}---MLP trained over Bad data with a window size of 0.5s. That each exemplar used a different architecture also suggests that the human subjects exhibited different strategies in performing the MARS task, to different effects.

Models trained over MARS data were found to also exemplify characteristics of different proficiencies when compared to participants in the VIP task, suggesting a level of generalizability between the two different environments. The selected architectures were then retrained using data from 3 VIP participants of each proficiency group to produce digital twins of VIP pilots. Table~\ref{tab:pilots} shows the performance characteristics of each pilot exemplar model. %The pilot evaluation was performed without any noise added to the actions. 
All other models that were trained over Good MARS data were reserved to act as candidate assistants, for a total of 21.\footnote{The selected Good pilot architecture was retrained with different weight initialization to create a distinct instance of the model to act as an assistant.}

%while the remaining models were removed from future evaluation.

% \subsubsection{imitation learning}

\vspace*{-2mm}
\section{Evaluation}
\label{sec:eval}

% \begin{figure}[t]
% \centering
% \includegraphics[width=0.98\columnwidth]{AuthorKit24/images/eval_pipeline_aaai24_paper.pdf} % Reduce the figure size so that it is slightly narrower than the column. Don't use precise values for figure width.This setup will avoid overfull boxes.
% \caption{PyVIP evaluation pipeline. \todo{Might need to move to appendix}}
% \label{pyvip-pipeline}
% \end{figure}

We perform two evaluations:
\begin{enumerate*}[label=\arabic*)]
    \item A high-throughput evaluation of pilot digital twins in co-performance of the VIP task with candidate assistants;
    \item A human subject study of human co-performance of the VIP task with the best-performing assistants from the digital twin study.
\end{enumerate*}

% Evaluation pairs a pilot model with a candidate assistant model and engages them in co-performance of the VIP task. Fig.~\ref{pyvip-pipeline} illustrates the pipeline with its 4 major components:
% \begin{enumerate*}[label=\arabic*)]
%     \item VIP
%     \item Crash predictor.
%     \item Assistant;
%     \item Pilot;
% \end{enumerate*}

The 4 major components of the evaluation pipeline include:
\begin{enumerate*}[label=\arabic*)]
    \item VIP
    \item Crash predictor;
    \item Assistant;
    \item Pilot.
\end{enumerate*}
The \textbf{VIP} component is {\it PyVIP}, a Python implementation for easy integration of ML models. 
The \textbf{Crash predictor} is a trained instance of the best crash prediction architecture reported in \citeauthor{wang2022crash} \shortcite{wang2022crash}: a stacked GRU trained over inputs like those in Sec.~\ref{sec:training} that predicts the likelihood of a crash occurring. Due to the crash predictor's high false positive rate, we added a \textit{crash probability threshold} of 0.8 where only highly imminent danger would permit assistant suggestions.\footnote{We follow~\citeauthor{wang2022crash} \shortcite{wang2022crash}'s hypothesis that too many false positives could cause a human pilot to lose trust in the assistant, but also need to not admit too many false negatives. 80\% represents a balance in these constraints.} The assistant would provide suggestions when either
\begin{enumerate*}[label=\arabic*)]
    \item crash probability is greater than the threshold {\it and} angular distance from the DOB exceeds 12$^\circ$, or
    \item angular distance from the DOB exceeds 15$^\circ$.
\end{enumerate*}
The \textbf{Assistant} observes task performance and makes suggestions when certain conditions are met.
The \textbf{Pilot} may be a digital twin or an actual human that controls the VIP. We used 6 digital twins---each of the architectures mentioned in Sec.~\ref{sssec:pilot-selection}, trained over both MARS and VIP data. Humans control the VIP with a joystick.
%\sm{
%\begin{itemize}
%    \item Without the crash predictor; the absolute angular position of the VIP from the DOB is outside than a preset \textit{SAFE ZONE}
%    \item With the crash predictor; either 1) crash probability is greater than the threshold and the above condition or 2) the position of the VIP from the DOB has entered a region considered the \textit{DANGER ZONE}
%\end{itemize}
%}

\vspace*{-2mm}
\subsection{Digital Twins Study}
\label{ssec:dt}
In these experiments, the pilot has an 80\% probability of accepting and executing an assistant suggestion instead of its own next action. If accepted, the pilot makes suggested deflections with a noise of $\mathcal{U}(-.05,.05)$ added to simulate human imprecision, after a $.4+\mathcal{U}(-.05,.05)s$ delay to simulate reaction time.\footnote{There is no prior work establishing the probability of human subjects following AI advice in \textit{this} task, but work on other tasks report $\sim$80\% correctness of/willingness to rely on AI advice \cite{du2021travelers,vodrahalli2022humans,liu2023using}. A 0.4$s$ reaction time is fast for an average human~\cite{nagler1973reaction}, but well slower than a trained pilot~\cite{binias2020prediction}.} %See appendix for rationales for these values and technical details.

We ran each evaluation for 3 30-sec. trials. Data was sampled at 200 Hz with each sample comprising of the angular position and velocity of the VIP, joystick deflection, crash probability, pilot and assistant's joystick deflections, which entity's deflection was performed, and whether the deflection made was destabilizing. 468 individual digital twin trials were collected, or 3.9 hours of data.

\subsubsection{Results}

\begin{table*}
\centering
 \resizebox{\textwidth}{!}{
 \begin{tabular}{lcccccc}
\toprule
\textbf{Assistant}        & \textbf{Crashes}$\downarrow$        & \textbf{\% destab.}$\downarrow$         & \textbf{$\mu\lvert\theta\rvert$ ($^\circ$)}$\downarrow$  & \textbf{$\sigma(\theta)$ ($^\circ$)}$\downarrow$   & \textbf{$\mu\lvert Mag\rvert_{vel}$ ($^\circ$/s)}$\downarrow$  &  \textbf{$vel$ RMS}$\downarrow$               \\
\midrule
\small SAC      & \small \begin{tabular}[c]{c} 0 / -5 / -25\\ -8 / -25 / -12\end{tabular}   & \small \begin{tabular}[c]{c}2.2 / 4.9 / -18.3\\ -31.8 / -38.3 / -28.5\end{tabular}  & \small \begin{tabular}[c]{c}0.4 / 1.2 / -3.6\\ -2.0 / -7.6 -1.0\end{tabular}   & \small \begin{tabular}[c]{c}-0.9 / -6.6 / -7.9\\ -1.9 / -6.7 / 7.9\end{tabular} & \small \begin{tabular}[c]{c}18.0 / -25.3 / -56.9\\ 27.0 / -37.7 / 7.5\end{tabular}   &  \small \begin{tabular}[c]{c}18.7 / -38.7 / -67.5\\ 24.4 / -42.2 / -6.9\end{tabular}   \\ \midrule

\small SAC-AIRL & \small \begin{tabular}[c]{c}-2 / -7 / -17\\ -15 / -33 / -12\end{tabular} & \small \begin{tabular}[c]{c}1.9 / -3.8 / -14.4\\ -36.8 / -41.5 / -28.5\end{tabular} & \small \begin{tabular}[c]{c}1.9 / -0.1 / 5.4\\ -0.5 / -6.1 / 0.9\end{tabular}  & \small \begin{tabular}[c]{c}0.9 / -7.7 / -0.3\\ -6.8 / -10.9 / 9.1\end{tabular} & \small \begin{tabular}[c]{c}-9.8 / -38.7 / -62.8\\ -22.5 / -66.6 / -8.7\end{tabular} &  \small \begin{tabular}[c]{c}-11.2 / -53.5 / -71.2\\ -29.9 / -72.9 / -20.3\end{tabular}  \\ \midrule

\small DDPG     & \small \begin{tabular}[c]{c}1 / -2 / -21\\ -11 / -23 / -12\end{tabular}  & \small \begin{tabular}[c]{c}2.3 / 3.7 / -17.3\\ -36.4 / -36.4 / -25.1\end{tabular}  & \small \begin{tabular}[c]{c}2.4 / -2.4 / -2.2\\ 2.6 / -7.5 / -1.0\end{tabular} & \small \begin{tabular}[c]{c}2.1 / -3.1 / -3.4\\ 3.5 / -3.6 / 9.2\end{tabular}   & \small \begin{tabular}[c]{c}25.3 / -11.0 / -43.1\\ 47.4 / -34.1 / 5.5\end{tabular}   & \small \begin{tabular}[c]{c}24.6 / -21.2 / -51.0\\ 43.5 / -38.6 / -12.5\end{tabular}   \\ \midrule

%\small MLP-G-0.5    & \small \begin{tabular}[c]{@{}c@{}}6 / 8 / -8\\ 8 / -20 / 8\end{tabular}        & \small \begin{tabular}[c]{@{}c@{}}11.2 / 14.2 / -1.6\\ -6.4 / -27.7 / -8.1\end{tabular}   & \small \begin{tabular}[c]{@{}c@{}}2.0 / 5.6 / 1.3\\ 0.3 / -4.4 / -0.2\end{tabular}   & \small \begin{tabular}[c]{@{}c@{}}0.1 / 0.9 / -3.7\\ 0.3 / -2.4 / 9.4\end{tabular}    & \small \begin{tabular}[c]{@{}c@{}}19.2 / -0.5 / -39.0\\ 39.2 / -32.5 / 39.6\end{tabular}   & \small \begin{tabular}[c]{@{}c@{}}26.0 / -5.8 / -40.2\\ 47.5 / -36.7 / 37.4\end{tabular}      \\ \midrule

\small MLP-GMB-0  & \small \begin{tabular}[c]{c}-2 / 4 / -11\\ -1 / -18 / -4\end{tabular}    & \small \begin{tabular}[c]{c}-0.5 / 7.0 / -8.8\\ -21.6 / -29.7 / -21.5\end{tabular}  & \small \begin{tabular}[c]{c}2.4 / 3.1 / 2.2\\ 2.1 / -6.8 / 4.0\end{tabular}    & \small \begin{tabular}[c]{c}2.2 / 2.8 / -1.7\\ 0.8 / -2.5 / 12.6\end{tabular}   & \small \begin{tabular}[c]{c}18.8 / 7.0 / -43.2\\ 31.4 / -15.5 / 12.2\end{tabular}    & \small \begin{tabular}[c]{c}24.5 / 1.3 / -48.0\\ 38.7 / -18.6 / 3.8\end{tabular}         \\ \midrule

\small LSTM-G-0.2   & \small \begin{tabular}[c]{c}3 / 14 / -7\\ 4 / -13 / 1\end{tabular}       & \small \begin{tabular}[c]{c}8.0 / 23.5 / -1.6\\ -11.0 / -20.5 / -8.8\end{tabular}   & \small \begin{tabular}[c]{c}3.6 / -0.2 / 0.9\\ 1.7 / -7.3 / 3.2\end{tabular}   & \small \begin{tabular}[c]{c}2.8 / 0.6 / 1.5\\ 2.7 / -4.0 / 12.6\end{tabular}    & \small \begin{tabular}[c]{c}25.4 / 15.6 / -33.7\\ 32.9 / -13.2 / 27.4\end{tabular}   & \small \begin{tabular}[c]{c}34.6 / 15.2 / -33.2\\ 41.5 / -14.2 / 24.8\end{tabular}     \\ %\midrule

%\small GRU-G-0.2    & \small \begin{tabular}[c]{@{}c@{}}7 / 6 / -8\\ 2 / -14 / 5\end{tabular}        & \small \begin{tabular}[c]{@{}c@{}}5.8 / 8.7 / -7.5\\ -15.3 / -24.9 / -11.9\end{tabular}   & \small \begin{tabular}[c]{@{}c@{}}3.1 / 4.0 / 0.1\\ -1.2 / -6.9 / 0.8\end{tabular}   & \small \begin{tabular}[c]{@{}c@{}}3.6 / 2.8 / 0.9\\ -0.5 / -3.1 / 10.6\end{tabular}   & \small \begin{tabular}[c]{@{}c@{}}25.9 / 5.5 / -32.2\\ 40.8 / -7.0 / 36.4\end{tabular}     & \small \begin{tabular}[c]{@{}c@{}}33.5 / -1.0 / -32.5\\ 58.3 / -7.8 / 30.8\end{tabular}        \\

\bottomrule
\end{tabular}}
\caption{Differences in performance with and without assistance (e.g., {\it 0} means no change in that metric, lower values are better---Sec.~\ref{ssec:perform}) In each cell, top line refers to MARS pilot models and bottom to VIP pilot models. Slashes separate Good/Medium/Bad pilot models.  Under Assistant, G/M/B denotes the proficiency of the assistant training data, decimals denote window size. Assistants shown achieved a significant reduction in at least one metric value. See appendix for results for all 26 assistants.}
\vspace*{-6mm}
\label{tab:diff-reduced}
\end{table*}

Table~\ref{tab:diff-reduced} shows performance differences between the digital twins when unaided and when aided by different assistants. Following the performance evaluation from Sec.~\ref{ssec:perform} (where lower metric values signal improvement), SAC-AIRL is the overall strongest assistant for digital twins, decreasing crashes, \% destabilizing deflections, and RMS velocity to a statistically significant level (all $p < 0.0001$ according to a paired two-tailed $t$-test). 

%\nk{2-29-24 Moved this to Section~\ref{ssec:perform}: A decrease in the metric usually means improvement, e.g., more time spent near the DOB, less oscillation, slower motion, smaller deflections, etc. However, the strongest metric---number of crashes---may be decreased significantly even if some other metrics rise.}

%Performance improvements with strong assistants are greater for less proficient pilot models. 
RL models are generally better assistants than DL models over human data.
%The best-performing recurrent models used a window size of 0.2s.
Interestingly, \textit{MLP-GMB-0} (MLP trained over all proficiencies with no window) decreased crashes as much as \textit{SAC-AIRL}, but for the Good MARS digital twin only.  The Medium exemplar architecture performs significantly worse when trained over VIP data than over MARS data, and the Bad VIP pilot performs much more like the Medium MARS pilot, while Good pilot models are roughly consistent with each other across tasks, suggesting that many strategies lead to poor task performance and relatively few do well. This also speaks to \textit{SAC-AIRL}'s ability to reduce crashes in both tasks for all proficiency levels. Other high-performing models also reduce crashes (\textit{DDPG}: $p=0.0002$, \textit{MLP-GMB-0}: $p = 0.0033$) and destabilizing deflections (\textit{DDPG}: $p = 0.0006$, \textit{MLP-GMB-0}: $p = 0.0080$) for digital twins trained over both MARS and VIP data, demonstrating transfer between digital twins trained over the different task data.

\vspace*{-1mm}
\subsection{Human Subject Study}
%\paragraph{Human-AI Co-Performance}
Results from the high-throughput digital twins setting indicated that the 5 assistant models shown in Table~\ref{tab:diff-reduced} had a statistically significant effect on one or more metrics in co-performance with digital twins trained over both MARS and VIP data. These models---3 RL-based models and 2 models trained over human data---were included as candidate assistants in the human subject study. This gave us a robust but tractable sample of assistants to assess in co-performance of the VIP task with real human subjects.

We recruited 20 healthy adult subjects (6 female, 13 male, 1 non-binary). Each subject participated in 2 experimental sessions separated by approximately one week. In {\bf Session 1}, subjects 
\begin{enumerate*}[label=\arabic*)]
    \item attempted to balance a 50\% coherent PyVIP RDK ($3\times30s$ trials);
    \item controlled the RDK with assistance from an AI model, rendered as left/right arrows indicating the direction of suggested deflection ($3\times30s$);
    \item watched the same AI control the RDK while providing directional suggestions via the joystick ($3\times30s$).
\end{enumerate*}
Participants were randomly assigned one of the candidate assistant models during Session 1---subjects were grouped into fours and each group received assistance from a single type of architecture. Subjects were not told which type of model they were receiving assistance from.

Between sessions, each assistant model was fine-tuned using data from Task 3 in Session 1. Episodes (consisting of input window and predicted action) where the direction of agent-predicted deflection conflicted with the direction of human deflection were stored. These human-in-the-loop (HITL) disagreement samples were used to fine-tune the model: the actor networks of the SAC and DDPG were fine-tuned using behavior cloning over the new data, the SAC-AIRL model was updated using AIRL over the new data, and the deep learning models underwent standard fine-tuning. %Hyperparameters are given in the appendix.

In {\bf Session 2}, subjects
\begin{enumerate*}[label=\arabic*)]
    \item undertook Task 1 as in Session 1 (solo RDK balancing---$3\times30s$);
    \item undertook AI-assisted balancing as in Session 1 Task 2 but with a different assistant model ($3\times30s$);
    \item undertook AI-assisted balancing with the version of their Session 2 Task 2 assistant fine-tuned with data from Session 1 subjects who interacted with that model type ($3\times30s$).
\end{enumerate*}
In Session 2 Task 2, subjects were assigned a non-fine-tuned model of a different architecture and in Session 2 Task 3 they were given an instance of that same model fine-tuned with HITL data from Session 1. Participants were not informed that the Session 2 Task 3 model was fine-tuned on Session 1 human data.

Finally, subjects took a survey, based on \citeauthor{muir1994trust} \shortcite{muir1994trust}, about their solo performance, how AI assistance changed their performance, and the level of trust they had in the assistant from each task.

\subsubsection{Results}

\begin{figure*}[h!]
\centering
\begin{subfigure}{.26\textwidth}
\centering
\caption{}
\includegraphics[width=0.98\columnwidth]{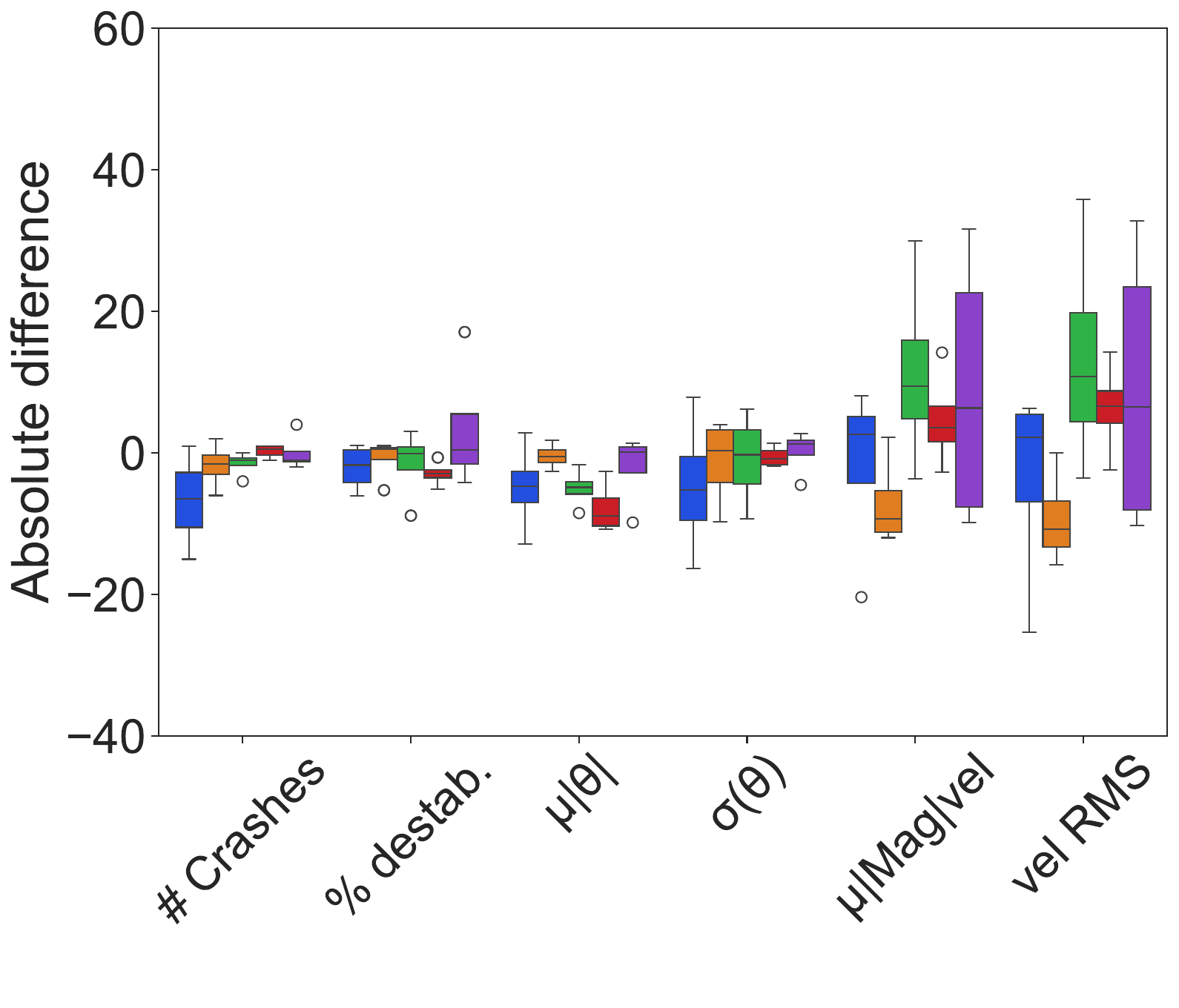}
\label{fig:human_ai_first_sess_abs_diff}
\end{subfigure}
\begin{subfigure}{.26\textwidth}
\centering
\caption{}
\includegraphics[width=0.98\columnwidth]{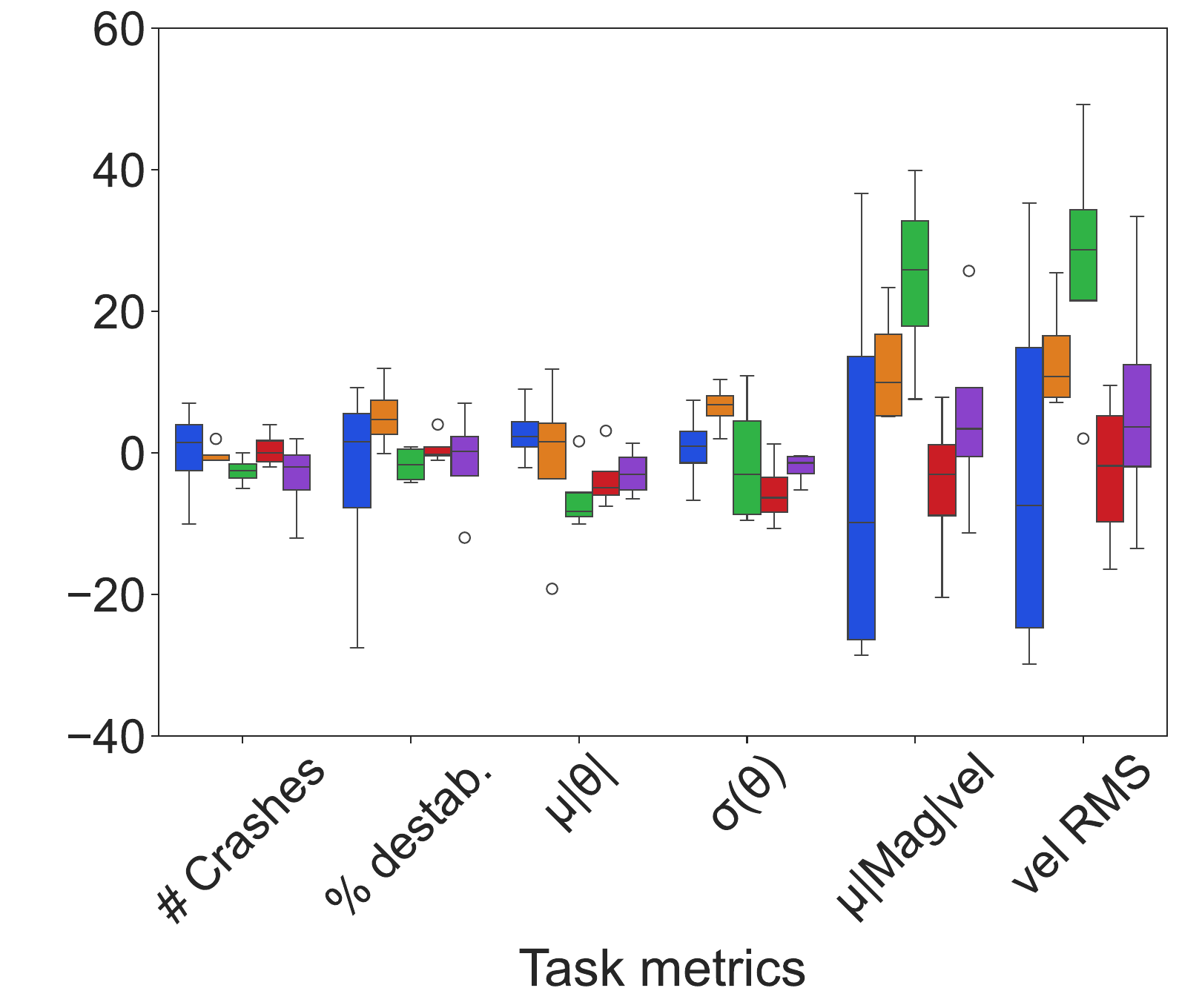}
\label{fig:human_ai_second_sess_orig_abs_diff}
\end{subfigure}
\begin{subfigure}{.35\textwidth}
\centering
\caption{}
\includegraphics[width=0.98\columnwidth]{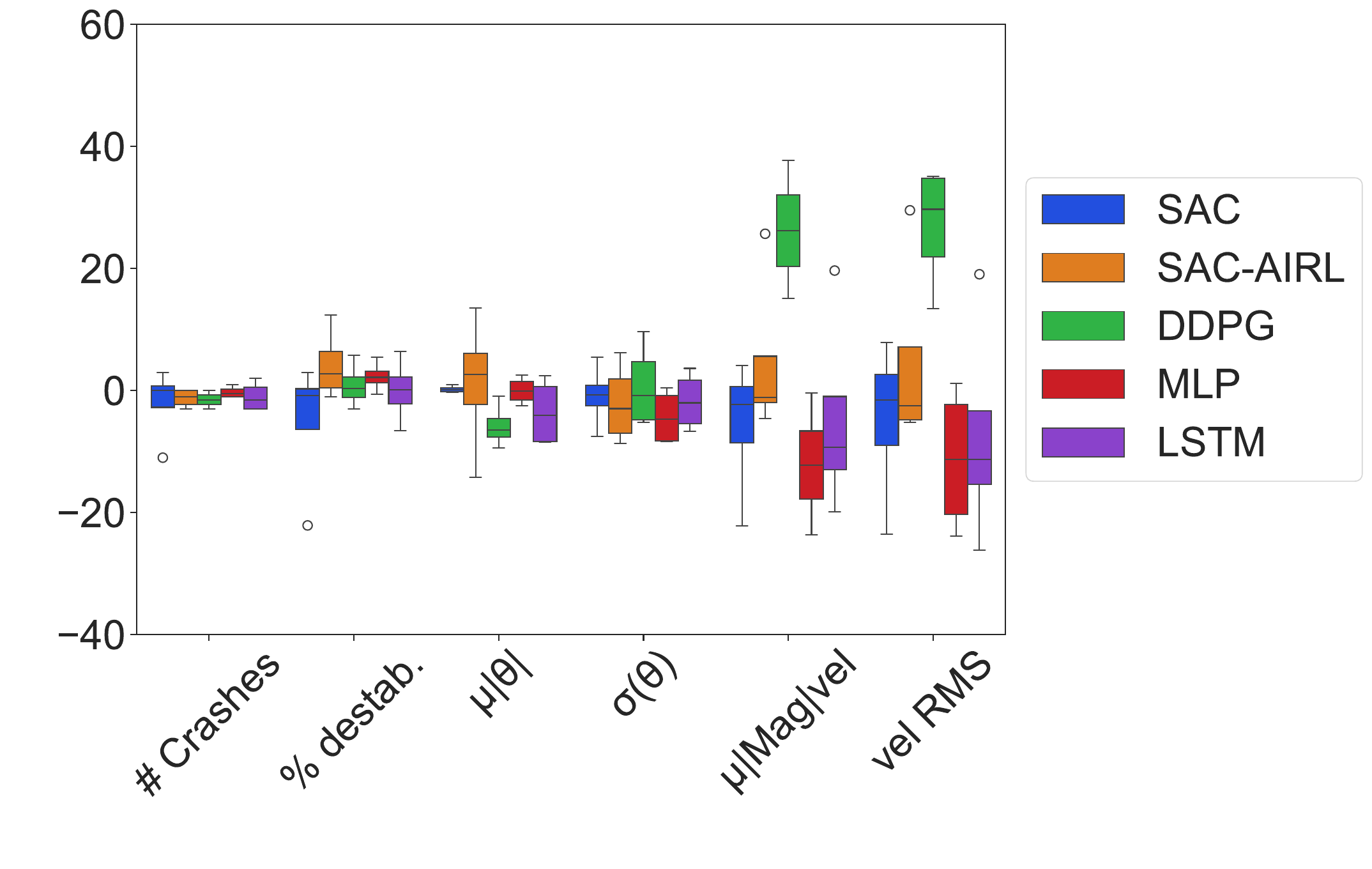}
\label{fig:human_ai_second_sess_retrained_abs_diff}
\end{subfigure}
\vspace*{-4mm}
\caption{Absolute differences between baseline human performance metrics compared to AI-assistance in (a) Session 1 Task 2, (b) Session 2 Task 2 (different assistant model), and (c) Session 2 Task 3 (fine-tuned Session 2 Task 2 assistant).}
\vspace*{-2mm}
\label{fig:human_ai_abs_diff}
\end{figure*}

We first assessed whether subjects displayed any adaptation to the balancing task within or across sessions that might confound apparent performance improvements due to AI assistance. Following \citeauthor{vimal2017learning} \cite{vimal2017learning}, in which participants in the disorienting MARS task showed minimal learning across consecutive days, we take performance in Session 1 Task 1 vs. Session 2 Task 1 (which were separated by approximately 1 week) as a baseline ``no learning'' condition in which participants lost familiarity with the task. We then compare performance differences between Session 1 Task 1 vs. Session 2 Task 1 and between Session 2 Task 1 vs. Session 2 Task 2. If the performance differences between Session 2 Task 1 and Session 2 Task 2 are similarly non-significant compared to the performance differences between Session 1 Task 1 and Session 2 Task 1, this indicates that no significant adaptation to the task occurred between tasks within Session 2, and likewise is unlikely to have occurred between Session 2 Task 2 and Session 2 Task 3; therefore apparent differences in Session 2 Task 2 and Session 2 Task 3 performance are likely to attributable to the nature of the AI assistance received.

We computed a score for each participant in each task of interest, given by Eq.~\ref{eq:objective-score},

\begin{equation}
                s=\big(\frac{60 - \mu\lvert\theta\rvert}{60}\big) + (1 - \frac{C}{90}) + (1 - \frac{pD}{100}) + \frac{pA}{100} + (\frac{R}{\text{max}_R}-\frac{C}{\text{max}_C}),
    \label{eq:objective-score}
\end{equation}

where $C$ is the count of crashes over the task ($3\times30s$ trials), $pD$ is the percentage of deflections that were destabilizing, $pA$ is the percentage of deflections that were anticipatory (see Fig.~\ref{fig:mars-vip}), $R$ is the task-level count of recoveries from beyond 20$^\circ$ away from the DOB to within 20$^\circ$ of the DOB, and $\text{max}_R$ and $\text{max}_C$ are the maximum number of recoveries and crashes in the data, respectively.

Because these scores are not normally distributed, we ran a Wilcoxon Signed-Rank test between Session 1 Task 1 and Session 2 Task 1 scores, and between Session 2 Task 1 and Session 2 Task 2 scores. No statistically significant differences were found between either pairing, with similar $p$-values (.2627 between Session 1 Task 1 and Session 2 Task 1, and .3681 between Session 2 Task 1 and Session 2 Task 2). This indicates that there was no adaptation to the task significant enough to confound performance differences attributable to AI assistance.

Fig.~\ref{fig:human_ai_abs_diff} shows the absolute difference in performance metrics between human solo VIP performance and 3 versions of AI-assisted performance for each model type: using the original model weights in Session 1 (\ref{fig:human_ai_first_sess_abs_diff}), the non-fine-tuned assistant from Session 2 (\ref{fig:human_ai_second_sess_orig_abs_diff}), and the Session 2 assistant fine-tuned on data from humans in Session 1 who interacted with the assistant of the same architecture (\ref{fig:human_ai_second_sess_retrained_abs_diff}). We see that although, like in the digital twin studies, the RL assistants in Session 1 were better at reducing the absolute number of crashes than DL assistants, this distinction often vanished or reversed after the assistant models were fine-tuned on participant data and then re-evaluated in Session 2. In Session 2 Task 3, the DL models fine-tuned on Session 1 data were often now better on average at reducing metrics associated with velocity and oscillation such as RMS velocity, velocity magnitude, and standard deviation of position. This effect is absent in Session 2 Task 2, where subjects were assisted by a different type of model.

%\vspace*{-1mm}
%\section{Results}

\vspace*{-2mm}
\section{Discussion}
\label{sec:disc}

% \begin{table}
% \centering
% \begin{tabular}{lcc}
% \toprule
%         {\bf Assistant} & {\bf $\mu$}  & {\bf $\sigma$}  \\
% \midrule
% \small SAC         & \small 69.17  & \small 50.47  \\
% \small SAC-AIRL    & \small 58.83  & \small 74.55  \\
% \small DDPG        & \small 132.17 & \small 164.25 \\  
% \small MLP-GMB-0   & \small 28.33  & \small 22.27  \\  
% \small LTSM-G-0.2  & \small 28.41  & \small 12.87  \\  
% \bottomrule
% \end{tabular}
% \caption{Mean and standard deviation of disagreement episodes logged during HITL study, by assistant model type.}
% \label{tab:disagreements}
% \end{table}

\begin{table}
\centering
\begin{tabular}{lccccc}
         & {\bf SAC} & {\bf SAC-AIRL} & {\bf DDPG} & {\bf MLP} & {\bf LSTM}  \\
\midrule
\small {\bf $\mu$}      & \small 69.17  & \small 58.83  & \small 132.17 & \small 28.33  & \small 28.41 \\
\small {\bf $\sigma$}   & \small 50.47  & \small 74.55  & \small 164.25 & \small 22.27  & \small 12.87 \\ 
\bottomrule
\end{tabular}
\caption{Mean \& SD of number of disagreement episodes logged during HITL study, by assistant model type.}
\vspace*{-8mm}
\label{tab:disagreements}
\end{table}

\begin{table*}[h!]
\centering
\begin{subtable}[t]{0.48\textwidth}
\centering
Perceived Performance Impact\\
\begin{tabular}{lcccccccc}
\toprule
        \multirow{2}{*}{\bf Assistant} & \multicolumn{3}{c}{\bf Task 2} && \multicolumn{3}{c}{\bf Task 3}  \\
        \cmidrule{2-4} \cmidrule{6-8} 
        & $+$ & $\sim$ & $-$ && $+$ & $\sim$ & $-$ \\
\midrule
\small Overall     & \small 50 & \small 15 & \small 35 && \small 55 & \small 30 & \small 15 \\
\small SAC         & \small 25 & \small 25 & \small 50 && \small 50 & \small 25 & \small 25 \\
\small SAC-AIRL    & \small 50 & \small 0  & \small 50 && \small 50 & \small 25 & \small 25 \\
\small DDPG        & \small 75 & \small 0  & \small 25 && \small 50 & \small 25 & \small 25 \\  
\small MLP-GMB-0   & \small 75 & \small 25 & \small 0  && \small 75 & \small 25 & \small 0   \\  
\small LSTM-G-0.2  & \small 25 & \small 25 & \small 50 && \small 50 & \small 50 & \small 0  \\  
\bottomrule
\end{tabular}
\caption{$+$: improved (incl. slightly/significantly), $\sim$: no change, $-$: decreased (incl. slightly/significantly).}
\label{tab:performance-impact}
\end{subtable}
\hfill
\begin{subtable}[t]{0.48\textwidth}
\centering
Reported Trust\\
\begin{tabular}{lcccccccc}
\toprule
        \multirow{2}{*}{\bf Assistant} & \multicolumn{3}{c}{\bf Task 2} && \multicolumn{3}{c}{\bf Task 3}  \\
        \cmidrule{2-4} \cmidrule{6-8} 
        & $+$ & $\sim$ & $-$ && $+$ & $\sim$ & $-$ \\
\midrule
\small Overall     & \small 25 & \small 35 & \small 40 && \small 30 & \small 45 & \small 25 \\
\small SAC         & \small 25 & \small 0  & \small 75 && \small 25 & \small 50 & \small 25 \\
\small SAC-AIRL    & \small 0  & \small 50 & \small 50 && \small 25 & \small 50 & \small 25 \\
\small DDPG        & \small 0  & \small 75 & \small 25 && \small 25 & \small 25 & \small 50 \\  
\small MLP-GMB-0   & \small 50 & \small 25 & \small 25 && \small 25 & \small 50 & \small 25 \\  
\small LSTM-G-0.2  & \small 50 & \small 25 & \small 25 && \small 50 & \small 50 & \small 0  \\  
\bottomrule
\end{tabular}
\caption{$+$: high to complete, $\sim$: moderate, $-$: low to none.}
\label{tab:trust}
\end{subtable}
\caption{Perceived performance impact of (\ref{tab:performance-impact}), and reported level of trust in (\ref{tab:trust}) Session 2 Task 2 and Task 3 assistants (as \%).}
\vspace*{-8mm}
\end{table*}

Like in the digital twins study, the SAC-AIRL assistant often helped the human subjects reduce crashes and oscillations. This is more pronounced in versions fine-tuned on HITL data, suggesting that models with a more human-like strategy contribute to this effect.

Assistance from the DDPG shows a strong tendency to increase RMS velocity and velocity magnitude values, and this is actually {\it more so} after HITL fine-tuning. This discrepancy is reflected in the number of disagreement episodes logged for each model type (Table~\ref{tab:disagreements}). Human subjects registered a much higher mean number of disagreements with the DDPG model---and RL models more generally---than with the DL models. This further indicates that the DDPG and RL models behave in ways that may contradict human intuition and/or physical instinct. Through their data and training, DL models are embodying the problem space and performing the task in a more human-like way, including transitioning from destabilizing to corrective and anticipatory deflections at distances from the DOB that align with human behaviors (cf. Fig.~\ref{fig:equiprobability}).

Due to different human reaction times, it is not possible to know exactly when human subjects followed assistant suggestions, but we can calculate a heuristic estimate based on instances where a subject deflects in the direction suggested by the AI within a threshold of the AI making a suggestion. Using a threshold of $450ms$, subjects followed AI suggestions approximately 44\% of the time ($\sigma=14\%$). This is significantly lower than the $\sim$80\% that can be seen in other domains (Sec.~\ref{ssec:dt}), suggesting particularities of the task need to be accounted for. The DDPG was the most followed assistant type at 53\%, followed closely by SAC-AIRL (51\%). The MLP was the least followed (32\%). Interestingly, fine-tuned assistants were followed $~$7\% {\it less} than non-fine-tuned assistants (4\% less when comparing Session 2 Task 3 to only Session 2 Task 2), even though subjects rated these assistants as more trusted and preferable (see below).

\paragraph{Trust survey}
Participants were asked to assess how the AI's suggestions changed their performance, and to rate their trust in the AI. We report survey results after Session 2, where participants also expressed their preference for one of the two assistant models used: an assistant with no specific fine-tuning ({\it Task 2} assistant), or one fine-tuned using Session 1 human data ({\it Task 3} assistant).

Table~\ref{tab:performance-impact} shows subjects' perception of their assistant's impact on their performance, overall and broken down by assistant type. Table~\ref{tab:trust} similarly shows the reported level of trust in each assistant. We see an overall trend toward better perceived impact on performance and more reported trust in the fine-tuned model when compared to the original, although interestingly some models to which participants ascribed a positive effect on performance (e.g., the MLP), were rated as less trusted after fine-tuning. At the assistant type level, these numbers should be taken in the context of small sample sizes ($N=4$). When asked to pick a preference between the Task 2 and Task 3 assistants, 15\% chose Task 2 and 70\% chose Task 3 (15\% no preference). %See appendix for analysis at a finer granularity.

\vspace*{-2mm}
\section{Conclusion and Future Work}
\label{sec:conc}

In this paper, we established a novel task of AI assistance in helping humans maintain balance in a disorienting condition. We first explored the space of possible AI assistance models using a high-throughput digital twins setting. The top performing models from this experiment were then used in a human-subject study to assess both performance impact and participants' attitudes toward different assistants.

Given certain data and training methods, AIs that were capable of performing IP balancing alone were also able to assist real humans in reducing crashes and oscillation. SAC-AIRL, which learns rewards implicit in human performance data, appeared to be an effective disorientation countermeasure in both the digital twins and human subject studies, by apparently embodying the problem space in a way that incorporates both physics and human signals. Although RL models on average make better assistants than DL models trained over human data, they do so by suggesting actions that often diverge significantly from the apparent model of the task captured in human actions. In the human subject studies: human subjects empirically perceived the RL models as performing the task incorrectly, and models that learned and embodied a human-like strategy through pretraining over human data, then were fine-tuned over more human data in the subject study, were able to significantly reduce factors related to oscillation and velocity.

%\todo{Moved from RW section: edit to fit}
\citeauthor{Palmer2016} \shortcite{Palmer2016} illustrate trust dimensions in the use of autonomous or automated systems, such as \textit{robustness} (handling perturbations/deviations appropriately), \textit{benevolence} (supporting mission and operator), and \textit{dynamism} (negotiating changes in environment). While our assistants' suggested actions may be appropriately corrective (benevolent), respond to pilot-induced perturbations such as ignoring cues (robust), and transfer between the MARS and VIP tasks (dynamic), they need to also be understandable in terms of the pilot's internal model of the situation to avoid corrections directly opposed to what the pilot expects (cf. the DDPG vs. MLP and LSTM in Fig.~\ref{fig:human_ai_second_sess_retrained_abs_diff}).  Our findings indicate that humans are in fact more receptive to assistance from an AI that demonstrates a more human-like, even if objectively suboptimal, balancing strategy.

Future study may investigate fine-tuning on data from a specific participant rather than an aggregate sample, to uncover person-specific patterns in task performance, or an investigation of modeling techniques that can account for the fact that human behavior is likely to change over time to account for assistance received from an AI agent, including one which is trained in real-time using live human feedback. Subsequent research may also involve transfer to more complicated conditions, like orientation in multiple roll planes or flight simulators, as well as investigating the transfer of AI assistance in the high-throughput VIP to the physical MARS.

There also remains the question of \textit{how} to deliver an AI assistant's cues to a human pilot. In this work we rendered visual indicators on the screen, but other modalities may include aurally rendered tones or vibrotactile cues (as in \cite{vimal2023vibrotactile}) to indicate the direction and magnitude of the corrective action, or linguistic instructions. In a previous experiment \citet{mannan-krishnaswamy-2022-go} presented evidence toward the utility of language understanding in task performance, and a multi-variable examination of intervention method and timing is another avenue of future study.

\vspace*{-2mm}
\begin{acks}
HND was supported by a Brandeis Summer Internship Grant.  VPV and PD were supported by Naval Aerospace Medical Research Unit Grant N0018923P0725. SM and NK would like to thank Sarath Sreedharan for his feedback on improving the framing and discussion of the paper. All authors would like to thank the many human subjects for contributing their valuable time toward the studies that led to this work, and the anonymous reviewers whose comments improved the final copy.
%To Robert, for the bagels and explaining CMYK and color spaces.
\end{acks}

\bibliographystyle{ACM-Reference-Format}
\bibliography{hai-24-mars-vip.bbl}

% \appendix









\end{document}